# What is online citizen science anyway? An educational perspective

CATHAL DOYLE, Victoria University of Wellington, New Zealand
YEVGENIYA LI, Victoria University of Wellington, New Zealand
MARKUS LUCZAK-ROESCH, Victoria University of Wellington, New Zealand
DAYLE ANDERSON, Victoria University of Wellington, New Zealand
BRIGITTE GLASSON, Victoria University of Wellington, New Zealand
MATTHEW BOUCHER, South Wellington Intermediate School, Wellington, New Zealand
CAROL BRIESEMAN, Hampton Hill School, Tawa, New Zealand
MELISSA COTON, Boulcott School, Lower Hutt, New Zealand
DIANNE CHRISTENSON, Koraunui School, Stokes Valley, New Zealand

In this paper we seek to contribute to the debate about the nature of citizen involvement in real scientific projects by the means of online tools that facilitate crowdsourcing and collaboration. We focus on an understudied area, the impact of online citizen science participation on the science education of school age children. We present a binary tree of online citizen science process flows and the results of an anonymous survey among primary school teachers in New Zealand that are known advocates of science education. Our findings reveal why teachers are interested in using online citizen science in classroom activities and what they are looking for when making their choice for a particular project to use. From these characteristics we derive recommendations for the optimal embedding of online citizen science in education related to the process, the context, and the dissemination of results.

## 1  INTRODUCTION

Participatory science, crowd-sourced science, or citizen science (CS), are a few of the many terms to refer to the modern phenomenon of scientific endeavour that involves amateur volunteers as contributors to real scientific projects. Some CS projects involve citizens - who are then also called citizen scientists - to collect field data such as sightings of specific plants and animals for example, while others ask the volunteers to look at some pre-collected data and annotate whether some particular features can be found within them (e.g. a particular pattern in a plotted graph). When these contributions are enabled by tools that rely on the Internet, researchers also use the term online citizen science (OCS). However, talking to different stakeholders who are involved in CS and OCS in practice or engaging with different research communities shows that there is still ambiguity in the terminology as well as the processes the different groups rely on.

In this paper we focus on OCS. A lot of research in this area has been done to understand the design of successful online citizen science projects from the perspective of the scientists who are tapping into citizen science as part of their projects, or from the perspective of facilitators of citizen science such as the various online citizen science platforms [1,2,3,4,5]. The questions addressed are often how one can ensure sustained motivation and engagement of citizens so that a project reaches its target [6]. However, this is a somewhat limited viewpoint, only regarding citizen science projects as a part of the research life-cycle of some professional scientists' line of inquiry. What is online citizen science anyway for the other groups that may engage with those projects? Do different groups have a different understanding of the purpose of online citizen science? May it be that professional scientists are developing past certain audiences interests and consequently limit the potential contribution online citizen science can make to the public understanding of science?

We contribute to the debate about the nature of online citizen science, focusing on an understudied area, the impact of online citizen science participation on the science education of children. We present the results of an anonymous survey among primary school teachers in New Zealand that are known advocates of science education. Our findings reveal that online citizen science is seen as an opportunity for students to engage with topics that otherwise are inaccessible in a classroom or lab setting (e.g. space) but that there is a mismatch between how online citizen science projects frame their suitability for in-class use (if they do that at all) and what teachers intend them to be used for. We also find that projects with a link to the local context are preferred by teachers, and highlight an important but understudied aspect that results from this, namely the cultural responsiveness of online citizen science.

The remainder of this paper is structured as follows: First, related work is outlined, which includes providing a definition for citizen science, and online citizen science. Based on this definition we propose an OCS process model from which we identify 16 process flows that demonstrate possible citizen scientists' involvement in different stages of an OCS. Afterwards we present the results of an anonymous survey involving primary school teachers who regularly teach science in New Zealand, followed by a discussion that synthesizes the results and links it with the related literature. We conclude the paper by summarizing the contributions and elaborating some future work stimulated by the research presented here.

## 2   FOUNDATIONS AND RELATED WORK

### 2.1   Science Education

Science is a key strategic focus of education in many countries in the world. In 2014, the New Zealand government, for example, launched a national strategic initiative entitled *A Nation of Curious Minds - He Whenua Hihiri i te Mahara*, aimed at improving public engagement with science and technology, and building greater scientific literacy amongst the New Zealand citizenry [7]. One key aim of this strategy is to enhance the role of education in improving engagement with science. The New Zealand Curriculum (NZC), introduced in 2007, already had at the heart of its science learning area an aim to build students' ability to engage with science and scientific issues of interest to them, through its *Participating and Contributing* sub-strand. The stated purpose of the science learning area as a whole is that "students explore how both the natural physical world and science itself work so that they can participate as critical, informed, and responsible citizens in a society in which science plays a significant role" [8, p.17]. This explicit embedding of science and inquiry-based learning in the NZC is just one example of how science education in general has become an integral part of many primary and secondary school curricula around the world [9,10]. Given this increased importance of science education in primary and secondary school curricula, it is noteworthy that evidence shows that student engagement with science is declining, reflecting international trends [11,12,13,14]. One way of improving scientific engagement may be through the inclusion of school programmes of citizen science which provides new opportunities for general public to participate in real scientific projects.

### 2.2   Citizen Science

Citizen science is a concept that has been practiced as far back as the 1700s [15], and is a scientific practice [15,16] that involves members of the public [17,18,19,20] actively engaging with professional scientists [16,18,20] in scientific work [18]. The members of the public participating can be referred to as citizen scientists [15,16] or volunteers [15,17,18,20].The engagement requires following an established protocol [18] that is created by the professional scientist [21], and can include one or more of the following tasks: data collection [15,17,20]; data processing [22,23]; data analysis & interpretation [15,17,18]; and/or dissemination of results [17,21]. Outcomes from citizen science projects include advancements in scientific research [16,19], as well as increasing the public's understanding of science [15,19,20]. Based on this understanding of citizen science, we adopt the following working definition: *Citizen science is a process that involves professional scientists and citizen scientists engaging on a scientific project. This engagement follows an established protocol, created by the professional scientists, which can include one or more of the following tasks: data collection, data processing, data analysis & interpretation, and/or dissemination of results. Outcomes should be advancements in scientific research, as well as an increase in the public's understanding of science.*

While this provides us with an understanding, and definition of CS, online citizen science (OCS) differs in various ways. We suggest that it is important to highlight these differences, in particular in the light of increasingly digital learning environments. Emphasising that digital technologies are a utility for some scientific endeavours - but not all of them (or not all stages of all of them) - bares the potential to create awareness to be critical when utilising almost ubiquitous technologies and ask whether the technologies are utilised purposefully. We also seek to stimulate a discussion about the digital traces that citizen scientists leave when engaging with OCS projects, which is an even bigger issue when the citizen scientists are students at schools. This, an overview of OCS is presented next, along with a working definition.

## 2.3   Online Citizen Science

Since the introduction of the Internet, and advancements in technology, citizen science has evolved to move the process online [15], allowing professional scientists to engage with citizen scientists in new ways [15,16,20]. This includes being able to provide easier access to large datasets [15,19]; making tools available to support engagement from citizen scientists that are geographically distributed [18,19,20]; enabling communication between citizen scientists [15]; and providing a wider reach to a broader audience of citizen scientists [18,20]. These citizen science projects can be aided by technology [18], or can be completely mediated online through technology [17,18]. This extension of citizen science has been called online citizen science [15,17,24], and digital citizen science [20]. Expanding on our working definition of citizen science we adopt the following working definition of online citizen science: *Online citizen science is an extension of citizen science, where the tasks to be completed are aided, or completely mediated, through the Internet. Engagement can occur in different ways such as providing larger datasets to be analysed; making tools available to support engagement from citizen scientists that are geographically distributed; enabling communication between citizen scientists; and providing a wider reach to a broader audience of citizen scientists.* Following this understanding and the working definition of OCS, we will now discuss a common process model that underpins OCS projects.

## 2.4   The online citizen science process

Bonney et al. [25] provide a process model for developing and implementing CS projects that can also be applied to OCS projects. The initial steps require choosing the scientific question(s) to be addressed; forming a project team; developing, testing, and refining protocols to be followed by the citizen scientists for the steps they'll be involved in; followed by recruiting and training the participants. While its argued that citizen scientists can participate in some of these steps (such as helping create the scientific question(s)) [26], it is much more common for professional scientists to complete these initial stages [25] where it is important to give careful consideration to each step to help ensure successful outcomes from the project [17]. Once completed, the next stages can involve data collection, data processing, data analysis & interpretation, and/or dissemination of results [15,17,21,23,25]. In terms of OCS, and depending how the professional scientists have set up the project, citizen scientists will participate in one to all of these process steps [17,22,27], as represented in Fig.1. This is to say, the professional scientist may already have the data collected but want the citizen scientists to analyse it, involving them in only one step, or they may wish for them to partake in the data gathering, and data analysis, involving them in two steps. Further to this, citizen scientists can play an active or passive role in these steps, depending on what is required of them when participating. Active participation involves citizen scientists using their cognitive skills and knowledge to complete a task, while passive participation occurs when citizen scientists share their computing resources through the internet to allow a project to run computationally intensive tasks [22].

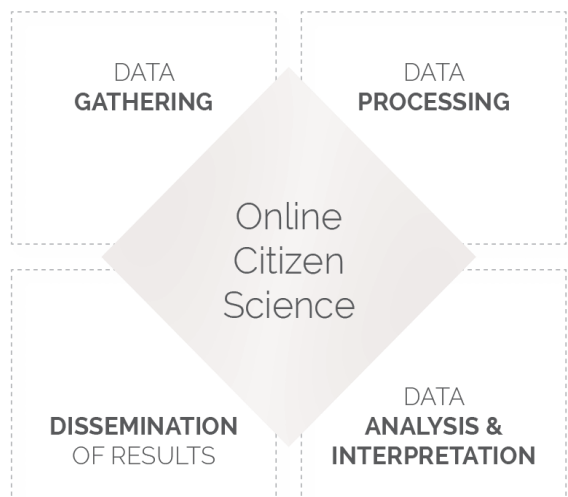

Fig.1. Online citizen science processes.

We will now provide a detailed explanation of the four steps from the perspective of the professional scientist from Fig.1, starting with data gathering.

*2.4.1   Data Gathering.* In terms of OCS, data gathering is an important step which can involve using citizen scientists as data collectors [15,17,21]. This is often done when the research question that is being asked requires the gathering of data that is on too large a scale for a professional scientist(s), or research team to achieve [21,28], and/or when it requires data to be gathered on a continental, or global basis [25,28]. Some things to consider when collecting data at such scale is to ensure the data is appropriate for the research question being asked [17,28]; that both the validity and the quality of the data is to the standards of the field [17,18,21]. To improve both the validity, and quality, of the data that is collected, professional scientists need to present clear and understandable data collection protocols to the citizen scientists [25]; ensure that the forms that will be used to gather the data are easy-to-use and logical [21,25,28]; and provide support for the citizen scientists if any issues arise when trying to understand the protocols, or submitting their data [25]. While it is possible that citizen scientists can play a passive role in data gathering [22], it is more common for them to have an active role.

*2.4.2   Data Processing*. There are two competing views when talking about data processing in relation to OCS. The first suggests that when citizen scientists are classifying, and/or annotating, gathered data, they are contributing to data processing for the project [18]. In contrast, Yadav and Darlington [22], drawing on Tinati et al. [1], propose that these tasks are part of data collection, and data processing is a different step. It consists of when data has been collected, the professional scientists might need to process it in order to be able to analyse it. To be able to do so, they may need computationally intensive processing procedures that they might not have access to [22,23]. It is here where citizen scientists can contribute their computing power by installing software that is provided, that the professional scientists can harness for the data processing [14,29]. As this study is focusing on OCS from a professional scientist's perspective, the latter view here is considered for data processing, as the former is a method of data collection. Further, while it is possible that citizen scientists can play an active role in data processing, it is more common for them to have passive roles [22].

*2.4.3   Data Analysis & Interpretation*. Data analysis and interpretation in OCS can involve using citizen scientists as data analysts [15,17,23,25]. It involves taking data that has already been gathered (and if necessary processed), and analysing, and interpreting, it to identify trends, patterns, and evidence-based findings [15,17,25,28] which can be done quickly [15]. There is a growing trend towards this step as larger datasets become available [24,30], which can be difficult for a professional scientist(s) or research team to analyse [15,28]. Some things to consider again are the provision of protocols on how citizen scientists should analyse the data [30]; ensuring the analysis meets the expectations of the field, while helping towards answering the research question(s) asked [17]; while making the data available, citizen scientists should be encouraged to manipulate and study the data [25] which can sometimes lead to serendipitous findings [10]. While it is possible that citizen scientists can play a passive role in data analysis and interpretation [22], it is more common for them to have an active role [23].

*2.4.4   Dissemination of Results*. The last stage, disseminating results, is used to inform citizen scientists of the project's progress, or results [21] and to publish the contributions to the scientific community [31]. Findings can be disseminated in different ways, such as conference or journal papers [15,17]; posted online [21]; presented at workshops [27] and forums/meetups [15]. While it may be seen as a reward to the citizen scientists for their participation [28], citizen scientists themselves can be involved in the dissemination process itself [15,17,27], such as being involved in the paper writing process [15,17]. However, while it is possible that citizen scientists can play an active role in disseminating results, it is more common for them to have a passive role. Following these explanations, a model is created next to represent 16 possible scientific process flows that can be applied in OCS projects.

# 3   UNPACKING THE OCS PROCESS FLOW

While there were several attempts to categorize OCS projects and develop CS/OCS domain-specific frameworks [22,27,32,33], the types of OCS projects based on the different degree of citizens involvement in activities along the scientific workflow remain unclear. One of the reasons is that scientific process workflows vary greatly among different disciplines. In this paper we adopt a simplified general linear scientific workflow, consisting of four broad stages: data collection, data processing, data analysis & interpretation, and dissemination of results. Using the four mentioned OCS process steps and adopting the logic of Zedtwitz et al. [34] approach, we arrive at a binary tree, as shown in Fig. 2, of 16 process flows that demonstrate possible citizen scientists' involvement in different stages of an OCS project. We denote professional scientists with "P" and citizen scientists with "C". The proposed model displays phases of the scientific workflow where citizen scientists participation is possible with an assumption that preliminary stages of an OCS project such as setting the research question(s), creating the protocol to be followed, and gathering the online citizen scientists to participate [25] have already been initiated by a professional scientist.

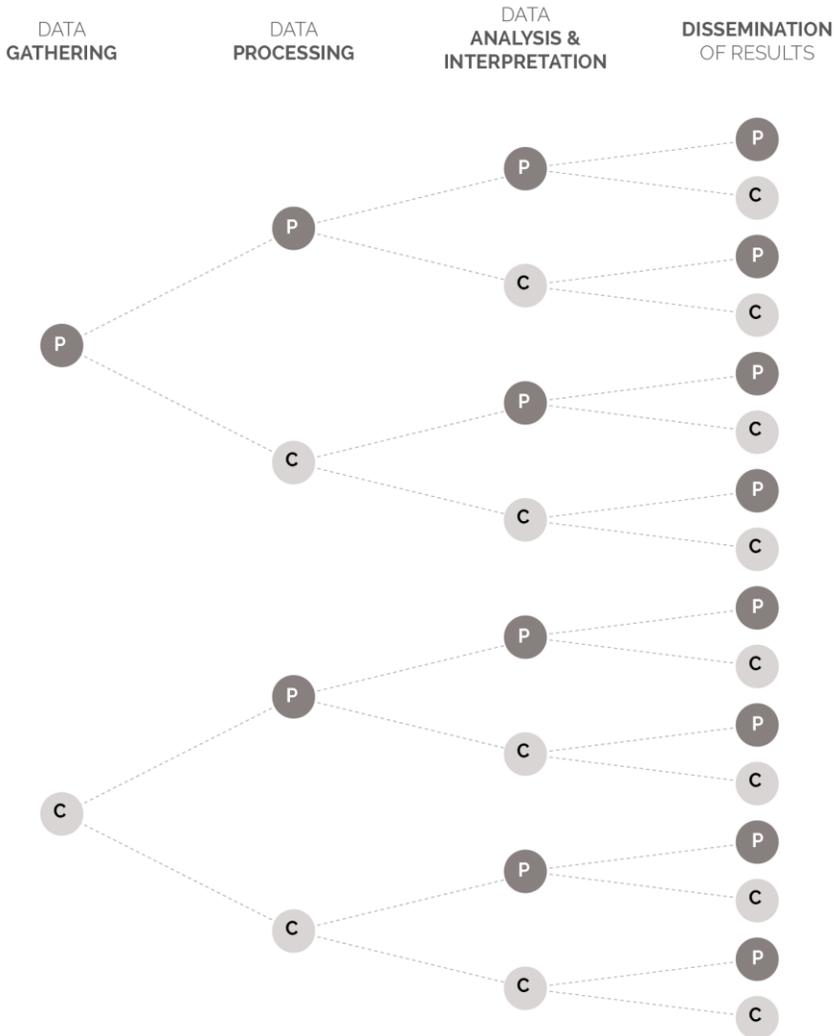

Fig. 2. A binary tree of online citizen science process flows.

The main advantage of the proposed model is that it clearly shows the variety of scientific projects where volunteer engagement may be totally absent (PPPP), limited to only one phase (e.g. CPPP or PCPP), extended to more than one phase (e.g. CCPP, CPPC) or presented in every single phase of the flow (CCCC). While the latter is not yet common and many OCS projects nowadays tend to involve community volunteers predominantly in the data collection stage (i.e. type CPPP), we believe that there is potential for OCS projects with more citizen scientists participation and in the following section we raise awareness about the lack of citizen involvement in stages like disseminating results and explain why such extensions are important both for scientists and community at large. The next section reports on the findings of an anonymous survey involving primary school teachers who regularly teach science in New Zealand.

## 4   RESEARCH DATA AND METHODS

In order to gather teachers' opinions about citizen science and online citizen science we have conducted an anonymous survey using a self-administered online questionnaire. The survey was distributed online using Qualtrics survey tools for a two-month period (December 2017 - February 2018). The survey consisted of ten mixed type questions (See Appendix A). While it was fully anonymous, all respondents were primary and secondary school teachers from the Science Teaching Leadership Programme (STLP) in New Zealand. Teachers' participation in the STLP programme means they have a sound understanding and experience in the science learning area of the New Zealand curriculum and an interest in promoting science among other teachers and schools. We first distributed the survey to a pilot group of 14 people, and after confirmation of the survey design, it was further distributed to the main group of 121 people (~75% female and 25% male respondents). In total, 42 respondents completed the survey, giving a response rate of 31%. The survey responses were read and independently coded by three researchers following an iterative coding approach [35]. Several main themes were identified, refined and grouped together into three large categories discussed in the following section.

## 5   RESULTS AND ANALYSIS

To begin with, this section summarises the teacher's knowledge of OCS. We found substantial differences between awareness of, and experience in, CS/OCS. While 29 out of 42 respondents (69%) confirmed that they knew what CS/OCS was, only 13 people (31%) had personal experience in using CS or OCS. Interestingly, among those 13, all but one have used OCS in the classroom. This indicates that respondents' engagement with CS/OCS was driven not only by self-interest but also by an intention to use OCS projects among children in their schools. Following this, we identified three themes, which reflect the results from the survey: (a) drivers for using OCS in the classroom; (b) criteria for choosing an OCS project; (c) challenges and opportunities. Each of these are discussed below, starting with drivers for using OCS in the classroom.

### 5.1   Drivers for using OCS in the classroom

There are several reasons why teachers decide to use OCS projects in the classroom. One of the main drivers is **children's involvement in, and contribution to real science**. Many respondents emphasized that OCS experience is inspiring for school children, who get excited with the idea that their efforts are part of real scientific projects and that their actions help to make a real difference.

> *"The students get ownership over their data collection and they know that they are making a difference to our place"*
> *"It is inspiring for the kids to feel that their Science work is authentic and part of a bigger picture - that their efforts will help in real life"*
> *"Kids could see meaning in what they were doing. Experiences were exciting hands on science."*

Another important rationale of OCS usage for schools was its **potential to interactively explain topics and matters that are not feasible or hard to present in lab or class experiments** (e.g. space and galaxies). Respondents noted a good fit of several OCS projects with their school curriculum and rich discussions held about nature of science stemming from OCS usage.

> *"It allowed the students to learn about galaxy types while participating in classifying them. It supported the skill of identifying types and the knowledge that galaxies have various shapes."*

Teachers could also see how particular OCS projects help to **develop students' science capabilities and knowledge building along with practical hands on experience at the same time**.

> *"I wanted students to get out and observe nature and upload data/pictures using devices they are familiar with so they can appreciate nature"*
> *"They had strong links to our kaupapa[1], they supported knowledge building but allowed for practical and hands on experiences as well"*

Teachers have used various OCS projects in the classroom including global (e.g. Galaxy Zoo, Zooniverse[2] projects) and local ones (e.g. Kereru project[3], Naturewatch NZ[4], Marine Metre Squared[5]). Further, several projects were conducted in cooperation with local government and councils (e.g. BOPRC projects[6], Wai Care projects[7], Pest Free Auckland[8]) which shows the nation's support for community-led collaboration between regional councils and local citizens for environment concerns, as well as potential for early science education and schools engagement in CS/OCS. The **importance of local-content** OCS projects was underscored by almost all of the survey respondents, which also corresponds to the overall trend in New Zealand OCS [36].

> *"Local matters."*
> *"Things that link to our local area and are tangible."*
> *"Local issues that will engage the children and allow them to make everyday connections with science issues."*

## 5.2   Criteria for choosing an OCS project

When asked about important aspects to consider when choosing a particular OCS project, participants named criteria such as: information about the project and its purpose, project timeframe and requirements, support provided, age level it is appropriate for, relevance to students' learning and local content. Many emphasized that the project should be easy to use while being fun and engaging for children.

> *"Clear, concise communication about the project that can be shared with the class"*
> *"Ease of interactiveness - ability to utilise it across all levels from Age 5 upward"*
> *"Relevance to topics and the skills students are using"*
> *"Something relevant and authentic to children, that is fun and they can see how they can make a difference."*
> *"Data collection processes that are not too onerous"*

One of the main findings of this survey was the **awareness that OCS engagement should not be limited to one stage** (e.g. data collection or classification of the given content) but should rather present a continuous process and include clear project purpose and description before citizens' participation, adequate support during engagement, and communication of results after the project finishes. While teachers originally chose projects in order to **support their curriculum and develop students' capability**, it is equally important for them to **be informed about further steps, know how their data is being used, and get notified about overall project outcomes**. The lack of feedback and follow-up information is common for many OCS projects. Professional scientists would usually provide information about the project, use citizen scientists to help with a particular task or research stage, and have the project outcome in the form of publications and research papers. These papers are often written by and for academics, are published in highly specialized journals and, thus, are often not accessible for general public. Our survey showed that citizen scientists (i.e. teachers and students) are keen to be not only practically involved in certain project stages but also to know the "bigger picture": how their local contribution added to a whole project and what are the final results:

> *"Feedback to students, updates on real progress, interactive discussion time, things to solve or challenge"*
> *"Feedback - it would be great to access maps, patterns, timelines to see what the data is showing locally and all over NZ or the world"*
> *"Clear feedback about ongoing trends, findings, or other ways in which the data is being used - in child speak"*
> *"A speaker or representative could present findings at school - this would reinforce the validity and worth of data gathering and OCS participation"*

---

[1] Philosophy and practice reflecting Māori cultural values.
[2] https://www.zooniverse.org/ Accessed on March 29, 2018
[3] http://www.projectkereru.org.nz/ Accessed on March 29, 2018
[4] http://naturewatch.org.nz/ Accessed on March 29, 2018
[5] https://www.mm2.net.nz/ Accessed on March 29, 2018
[6] https://projects.boprc.govt.nz/ Accessed on March 29, 2018
[7] https://waicare.org.nz/ Accessed on March 29, 2018
[8] https://goo.gl/xp1vgs / Accessed on March 29, 2018

In the context of school children feedback, and clear communication of results, **allow students to see the process of the science investigation**. Some respondents even advocate for long-going OCS projects where data is accumulated *"so tracking can take place over the years and then be followed up at high school"*. Children engaged in OCS projects are important for dissemination of scientific knowledge and outcomes as they share information with their parents and friends and propagate this knowledge in the future thus increasing scientific literacy far wider than themselves and their classrooms.

## 5.3   Challenges and opportunities for using OCS in the classroom

Interestingly, respondents also raised awareness that currently there is a variety of local OCS projects related to nature and biology while a **lack of OCS projects with physics, chemistry and geology content**. This corresponds to a worldwide trend. For example, in the case of Zooniverse - one of the largest OCS platforms - there are more than 50 projects in Biology and Nature, while currently only eight under the "Physics" category. This trend, on the one hand, indicates that certain fields and disciplines might be more suitable for CS and OCS and are better adapted for volunteers and community involvement based on the needs and requirements of the project (e.g. large amount of data, spread physical locations, easier access for local citizens to particular areas, etc.). It is also supported by previous literature on the criteria and decision framework whether citizen science is suitable for a particular project [32]. On the other hand, this trend provides new opportunities and potential for OCS growth in the mentioned disciplines:

> *"Lots of biological science projects available and gathering data but would like to see some NZ content in physics and chemistry as well as geology."*

Yet another challenge is that **most OCS projects are not originally intended for use in the classroom**. OCS projects usually do not have strict requirements for volunteers' qualifications and are open to contributors of various backgrounds, age, skills and prior OCS experience. While the absence of such criteria encourages and promotes wider citizen participation at large, it also means that professional scientists often do not clearly state which knowledge, capabilities, level of computer skills, etc. might be necessary to fulfil the tasks and do not take into consideration the potential usage of their projects in schools.

> *"Data entry can be awkward at times"*
> *"It was too much work to gather the data for this age group"*

We should note, however, that some OCS projects do acknowledge this potential and provide additional educational and supporting materials for running their project in the classroom (e.g. Sunspotter project[9], Planet Four[10]). A discussion of these findings is had next.

## 6    DISCUSSION

### 6.1    OCS Process Flow for Education

The OCS literature indicates that there are two outcomes expected from participating in an OCS project for the citizen scientists involved: first they gain a better understanding of the scientific process and realisation of how science is conducted; second is to learn about the actual topic that the OCS project is focusing on [15,19]. The findings from this study indicate that both of these outcomes are also important aspects as to why teachers would be willing to adopt OCS projects for their classrooms. Further, while citizen scientists can be active, or passive, in terms of their involvement in an OCS project, teachers indicated that for students to get the benefits to their learning, it is more valuable that students be involved in active stages of the OCS process. In doing so, students are getting "hands on" experience of the process(es) they are involved in, such as understanding how data is gathered, and/or how data can be analysed and interpreted, for a real world scientific project, while also learning about the topic of the project. This would suggest that to encourage both outcomes from an OCS project, the desirable OCS process flow that can be followed is CPCC as shown in Fig. 3. Such a flow would indicate that students are actively involved in the data collection step of an OCS project; the professional scientist may then take that data and process it in some way - this is not as important to students as it often represents a passive activity. While it might be argued that there is still learning here, it is not something that the teachers indicated as being important. Once the data has been processed, the students can become involved again, where they are active again in analysing & interpreting the data. Lastly, the stage for disseminating results is important enough to be discussed in detail below.

---

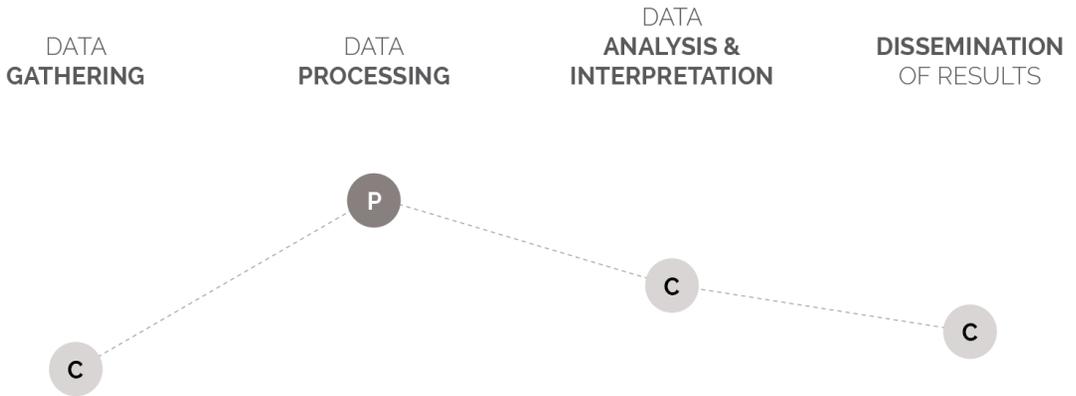

Fig. 3. Desirable OCS Process Flow for Education.

It's also understood from the survey that engagement between the professional and citizen scientists is important, and should not be limited to one stage (e.g. data collection). Instead it should be viewed as a continuous process where a clear project purpose and description must be provided before citizen scientists participate, providing a clear project context for the teachers; support is provided during the stages that the citizen scientists engage with the project; and communication of results after the project finishes is important for learning. These are discussed in more detail in the following sections.

## 6.2   Project context

When setting up an OCS project, there are many aspects that need to be considered such as choosing the research question(s) to be addressed; forming a project team (if necessary); and developing, testing, and refining protocols to be followed by the citizen scientists for the steps they'll be involved in. While some argue these are stages that citizen scientists can be involved in, the data highlights they are not stages that teachers are particularly interested in. This could be due to time constraints on trying to find projects that fit into their already busy teaching plans; a lack of experience from the students to be able to consider how such aspects can be constructed due to their age, knowledge and/or this being the first OCS project they are involved in. Instead it is easier for the professional scientist(s) to spend the time in setting up these aspects, creating relevant research questions based on their experience and knowledge of what they want to study.

However, what teachers are interested in, is the context around OCS projects, which includes some of the information that professional scientists need to provide in the initial steps of an OCS project. This helps teachers to understand the requirements of a project, if it is suitable for the students, and how it can be used in their classrooms. What wasn't so evident before, is what this context should include, so presented in Table 1 is a list that teachers stated as important for them when considering an OCS project to adopt, and an explanation of each. Each of these are important elements that professional scientists should consider when setting up an OCS project, and make it explicitly available so teachers can get a good understanding of the project, and its relevance to their students, and topics that they are teaching.

**Table 1. Important contextual description required for an OCS project**

| Context | Explanation |
| --- | --- |
| Purpose of the project | Provide a general overview of the project, explaining the purpose, and what it wants to achieve, including the area of focus |
| Project timeframe | State when the project will start, how long it will run for, what stages participants will be required to complete their tasks (including when they will be expected to do so), and when and how results will be disseminated |
| Requirements | Technical requirements such as<br> - Devices that can be used<br> - App(s) and software needed<br> - Accessibility (is an individual account required, or will a group one do?)<br>Skill requirements such as<br> - Numeracy skills<br> - Literacy skills<br> - Technical/Computer skills |
| Support provided | Specify what support will be provided for the participants for the different stages of the project. This includes materials that are available to gain an understanding of the project, what's expected of the participants, , such as videos and/or documentation, |
| Age level | Indicate what the minimum suitable age level for participation is |
| Relevance to students' learning | Explain the area of focus for the project, what aspects of this area will be touched on during the project, and how students will be exposed to it by their participation |
| Geographic scope | Indicate if the OCS project is of local, national, or international, interest |
| Data availability | Indicate if the data that is gathered, or used, in the project is available for access outside of the project itself for further manipulation, such as being able to collaboratively compare it with other classes/schools locally, nationally, and/or internationally |

It's also worth noting the importance for professional scientists to properly prepare these initial stages carefully, as failure to do so can result in failed projects [17]. If the stages are properly set up, it can lead to participants being more engaged with the project, improving the chances of project success. For example, when creating protocols to train participants for the task(s) that they are going to complete, it may suffice to provide written documentation that can be followed, or videos may be necessary to show the participants what they must do. When teachers are reviewing OCS projects for their classrooms, they review these protocols to ensure that they are sufficient for the students in their class, so in creating them, the professional scientists should give consideration on who the participants will be, what level of detail is required, and how best to convey this detail. Further, professional scientists should also consider the UX and UI design, the ease of use of the platform they are providing, how it can be accessed and used, and whether they can make it fun and engaging. By making such information explicit, it allows teachers, and general members of the public, to better understand the project as a whole, what's expected of them, and allows them to make a more informed decision on whether to participate in the OCS project or not.

## 6.3   Dissemination of results: Beyond the academic paper

Evident from the results of the survey is that dissemination of results is a critical aspect. While developing science capabilities and a better understanding of the scientific process are primary reasons for OCS involvement, teachers indicated the importance of seeing the results of contributions made. They indicated the importance of being able to see the output(s) of the OCS project that the students participated in, to see how their efforts contributed to a real world scientific project, and that their active involvement had some tangible outcomes. This is to say, they don't just want to be "used" so data can be gathered, or analysed, but they want to feel like they have helped make a difference, no matter how big in relation to other aspects of the real scientific life-cycle. Such feedback is expected to provide satisfaction for the work done, and stimulate a willingness to engage in OCS projects in the future.

However, this is in contrast to the typical goal of professional scientists. Often, professional scientists are only concerned with the positive findings they are aiming for in relation to their research question(s), and the publication(s) that will result from those positive findings. Such a view is taken as publications are the currency required for promotion and scholarly recognition in their respective fields. This is the same for professional scientists working on OCS projects, but it can be argued that such a view prevents some of the goals that OCS is supposed to help achieve, and possibly hinders the growth of science that can be achieved through such projects.

Another issue is that the results of the project may not be easily accessible by general public if they are presented in specialized conferences and academic journals. Scientific papers written by and for academics might also be difficult for understanding as the results are presented using specialist language and formatting. The reading age requirements to understand these papers is often beyond the scope of primary students.

Therefore, it is important that if professional scientists are engaging with students, and we could argue, the general public, in their OCS projects, they need to be prepared to look beyond the academic paper. They need to engage with the students and the public who are assisting them, disseminating the results in ways that they can interpret and understand. This can be achieved in numerous forms, such as presentations, student and non-scientist friendly blog posts, wikis, Q&A sessions with the students, and forums. Doing so can help students and the public see the "bigger" picture in terms of their contributions, and helps form their opinion on science, and professional scientists themselves. This can encourage the attitude of "I can do science" [15, p.6]. Further, this also ties in with Raddick et al. [15] argument, which draws on Chambers [29], that students and the public become more aware of scientists, how they work, and what they look like, discarding the image of white old men who work alone, giving them confidence that they can take on such roles in the future. Finally, students themselves become disseminators of knowledge, where they are bringing their new knowledge back to the home, and community in which they are part of. Also, in line with the goal of curious minds [7], it is not just about trying to create a nation of scientists, but encourage citizens to be able to use their science knowledge to participate as critical, informed, and responsible citizens.

## 6.4   Towards culturally inclusive online citizen science

Our study was carried out in New Zealand. While there is a significant overlap between the New Zealand curriculum to the curricula of other countries, New Zealand features a unique bi-cultural context. Hence, cultural responsiveness of the online citizen science projects that are to be selected for learning activities is a major concern to teachers and students alike. To our best knowledge we are not aware of a single online citizen science project that addresses a global audience, and incorporates any form of cultural responsiveness. This could start from being available in other languages than just English, but could go as far as a careful consideration of what may be regarded sacred topics in indigenous cultures, or even to questioning the nature of the scientific inquiry and the adequacy of the underlying knowledge system [37].

While this seems to be a major cultural shortcoming of the existing online citizen science projects, it can equally be a great opportunity for a large scale participatory exercise. Teachers can pick up the topics represented by the different OCS projects and ask their students to talk to senior citizens of their indigenous communities to provide their stories about these topics, e.g. in New Zealand the Matariki[11] and navigation stories about the stars at night. This would not only be a genuine enrichment of the citizen science landscape but would contribute to the importance of acknowledging this knowledge as being owned by the indigenous people of an area. This further contributes to valuing citizens for their many knowledge systems and validating their place in citizen science.

---

[11] Māori name for the cluster of stars also known as the Pleiades

## 7    CONCLUSION

In this research we answered one central question: What is online citizen science anyway? We addressed this question from the perspective of science teaching and learning in conformance with a real science education curriculum. This makes our work substantially different from previous studies that investigated learning in the context of online citizen science but were limited to informal learning of any kind of volunteer participants [38].

This paper contributes to the knowledge base of the computer supported collaborative work research community by tapping into an understudied area - online citizen science in the context of formal primary and secondary school education. Our findings are a significant step towards an understanding of what teachers expect and need from an online citizen science project to be suitable for their teaching. Our paper also makes a substantial contribution to online citizen science platform providers' understanding of how they should frame their projects to be of use in real science education, which goes beyond the provisioning of some pre-designed teaching cases as these may miss the target of the specific requirements of national curricula and teachers' needs. The teacher survey also provides some initial insights into UX and UI design considerations OCS platform providers may need to be aware of when setting up projects with school children in real educational settings in mind.

We undertook an initial decomposition of the OCS process to highlight the difference between citizen and professional scientist involvement in the different stages, and to indicate which process might be preferred by teachers based on our survey results. We acknowledge that even this decomposition fails to account for the real complexity of the scientific process. In a follow on study we will dig deeper into the iterative and intertwined nature of the scientific process by assessing the various sequences of data collection, data processing, data analysis and interpretation, and dissemination of results we can find in a broad range of actual online citizen science projects. This will complement the results described here, which brought in the teachers' perspective, and outlined what is desired in a formal educational context. The future work will also focus more on the proposed model of OCS flows (Fig. 2) with more detailed analysis of features, opportunities and challenges of each of the 16 OCS process flows.

**APPENDIX A**

<div align="center">Survey questions</div>

1. Do you know what Online Citizen Science (OCS) is?
   a. Yes
   b. No [if respondent chose "No" a short explanation and example was provided]
2. Do you have experience in using Citizen Science or Online Citizen Science personally?
   a. Yes
   b. No
3. [Only shown if answer was "Yes" in Q2] Please describe briefly any project(s) you have been personally involved in.
4. [Only shown if answer was "Yes" in Q1] Do you have experience in using Online Citizen Science in the classroom?
   a. Yes
   b. No
5. [Only shown if answer was "Yes" in Q4] Which project(s) did you use?
6. [Only shown if answer was "Yes" in Q4] What made you choose that project(s)?
7. [Only shown if answer was "Yes" in Q4] What did you like about that experience?
8. [Only shown if answer was "Yes" in Q4] What could be improved?
9. Thinking about your school's Science Programme, what would you look for in an Online Citizen Science project?
10. What information would you like to know about an Online Citizen Science project before you decide to use it in class?